\documentclass[final,5p,times,twocolumn]{elsarticle}

\usepackage{multicol}
\usepackage{color}
\usepackage{xspace}
\usepackage{hyperref}

\usepackage{amssymb}
\journal{Physics Letters B}

\begin{document}

\begin{frontmatter}

%% Title, authors and addresses

%% use the tnoteref command within \title for footnotes;
%% use the tnotetext command for theassociated footnote;
%% use the fnref command within \author or \address for footnotes;
%% use the fntext command for theassociated footnote;
%% use the corref command within \author for corresponding author footnotes;
%% use the cortext command for theassociated footnote;
%% use the ead command for the email address,
%% and the form \ead[url] for the home page:
%% \title{Title\tnoteref{label1}}
%% \tnotetext[label1]{}
%% \author{Name\corref{cor1}\fnref{label2}}
%% \ead{email address}
%% \ead[url]{home page}
%% \fntext[label2]{}
%% \cortext[cor1]{}
%% \address{Address\fnref{label3}}
%% \fntext[label3]{}

\title{Observation of a first-order pairing phase transition in atomic nuclei}

%% use optional labels to link authors explicitly to addresses:
%% \author[label1,label2]{}
%% \address[label1]{}
%% \address[label2]{}

\author[label1,label2]{L.~G.~Moretto}
\ead{lgmoretto@lbl.gov}
\address[label1]{Department of Chemistry, University of California, Berkeley, CA 94720, USA}
\address[label2]{Lawrence Berkeley National Laboratory, 1 Cyclotron Road, Berkeley, CA 94720, USA}

\author[label3]{A.~C.~Larsen}
\ead{a.c.larsen@fys.uio.no}
\address[label3]{Department of Physics, University of Oslo, N-0316 Oslo, Norway}

\author[label3]{F.~Giacoppo}
\author[label3]{M.~Guttormsen}
\author[label3]{S.~Siem}
\author[label4]{A.~V.~Voinov}
\address[label4]{Department of Physics and Astronomy, Ohio University, Athens, Ohio 45701, USA}

\date{\today}

\begin{abstract}
Experimental nuclear level densities at excitation energies below the neutron threshold follow closely 
a constant-temperature shape. This dependence is unexpected and poorly understood. 
In this work, a fundamental explanation of the observed constant-temperature behavior in atomic nuclei 
is presented for the first time. It is shown that the experimental data portray a 
\textit{first-order} phase transition from a superfluid to an ideal gas of non-interacting
quasiparticles. Even-even, odd-$A$, and odd-odd level densities show in detail the behavior of gap- and 
gapless superconductors also observed in solid-state physics. 
These results and analysis should find a direct application to mesoscopic systems such as superconducting clusters.

\end{abstract}

\begin{keyword}
%% keywords here, in the form: keyword \sep keyword
Experimental nuclear level densities \sep Pairing \sep Constant temperature
\sep First-order phase transition

%% PACS codes here, in the form: \PACS code \sep code
\PACS 21.10.Ma \sep 21.10.-k \sep 25.40.Hs
% PACS Numbers: 21.10.Ma (level density), 
%               21.10.-k (Properties of nuclei; nuclear energy levels), 
%               21.60.Jz (HF & QRPA)
% 		25.20.Lj (Photoproduction reactions)
%		24.30.Gd Other resonances
%		25.55.Hp	3He transfer reactions
%		27.40.+z	39 ² A ² 58
%		27.50.+e	59 ² A ² 89
%		27.60.+j	90 ² A ² 149
%		25.40.Hs	Transfer reactions, nucleon-induced
% From Voinov's PRL: 25.40.Lw, 25.20.Lj, 25.55.Hp, 27.40.+z

%% MSC codes here, in the form: \MSC code \sep code
%% or \MSC[2008] code \sep code (2000 is the default)

\end{keyword}

\end{frontmatter}

%% \linenumbers

%%%%%%%%%%%%%%%%%%%%%
% INTRO/MOTIVATION  %
%%%%%%%%%%%%%%%%%%%%%
Phase transitions and phase coexistence are phenomena that are eagerly sought after and investigated
both experimentally and theoretically, since they involve dramatic changes in the thermodynamic properties
of the system as it undergoes the transition from one phase to another. In hadronic systems the quark 
confinement-deconfinement transition has been under vigorous scrutiny and intense experimental investigation.
In nuclear systems, the liquid-to-vapor phase transition has long been searched for and finally 
found~\cite{moretto2011,elliot2013}.

For atomic nuclei away from closed shells and at low excitation energies,
the pairing force is the dominant two-body residual interaction and plays a major role in defining their properties 
together with the
one-body shell-model component (the single-particle term)~\cite{dean2003}. 
The effect of pairing on the single-particle levels is the presence of a gap and the compression of the
quasiparticle spectrum compared to the one-body single-particle spectrum.

For conventional superconductors, 
the standard Bardeen-Cooper-Schrieffer (BCS) pairing description~\cite{BCS1957} predicts a critical temperature/angular momentum
at which the superconducting phase reverts to the normal one through a second-order phase transition. However,
for atomic nuclei, this second-order phase transition has never been truly verified experimentally, 
in spite of long and intense efforts. Within the canonical ensemble theory, attempts have been made to assign 
a second-order pairing phase transition to the $S$-shape of the canonical heat capacity~\cite{schiller2001}. 
This approach is questionable in view of the special case of constant temperature, 
as discussed farther down and also emphasized in recent works~\cite{nyhus2012,guttormsen2013}.
On the other hand, first-order phase transitions can arise from the BCS 
Hamiltonian under specific circumstances, as demonstrated in Ref.~\cite{moretto1975}. 

In this Letter, we show that a first-order, rather than a second-order, phase transition is dramatically evident in
experimental nuclear level densities below neutron threshold, that this transition is endowed with constant temperature and latent heat,
that we can identify the two phases in equilibrium and their nature,
and that this first-order transition is, in most nuclei,
indisputably related to the presence of an energy gap in the quasiparticle spectrum. 

A large body of high-quality nuclear level-density data are now 
available in the literature~\cite{ocl-website,alexander-website,voinov2009}. The stunning, common feature of the
level densities, particularly evident for deformed, midshell nuclei, is the linear dependence of their logarithm with excitation 
energy. Above $\approx 2\Delta_0$, where $\Delta_0$ is the pair-gap parameter, and up to about 
the neutron separation energy, they are well described by the constant-temperature expression 
proposed by Ericson~\cite{ericson1959}, and Gilbert and Cameron~\cite{GC1965}:
\begin{equation}
\rho(E) \propto \exp(E/T),
\label{eq:ct_nld}
\end{equation}
where $E$ is the excitation energy and $T$ is the constant nuclear temperature. They found this 
expression to be in good agreement with the cumulative number of levels at low excitation energy, but did not
provide any fundamental, quantitative explanation for this relation. Moreover, the constant-temperature expression 
is in striking 
contrast to the expected Fermi-gas behavior as 
first outlined by Bethe~\cite{bethe1936}, 
predicting a square-root dependence of the level density
with excitation energy:
\begin{equation}
\rho(E) \propto \exp(2\sqrt{aE}), 
\label{eq:fermi_nld}
\end{equation}
where $a$ is the level-density parameter. 
%---------------------------------------------------%
 \begin{figure*}[hbt]
 \begin{center}
 \includegraphics[clip,width=2.0\columnwidth]{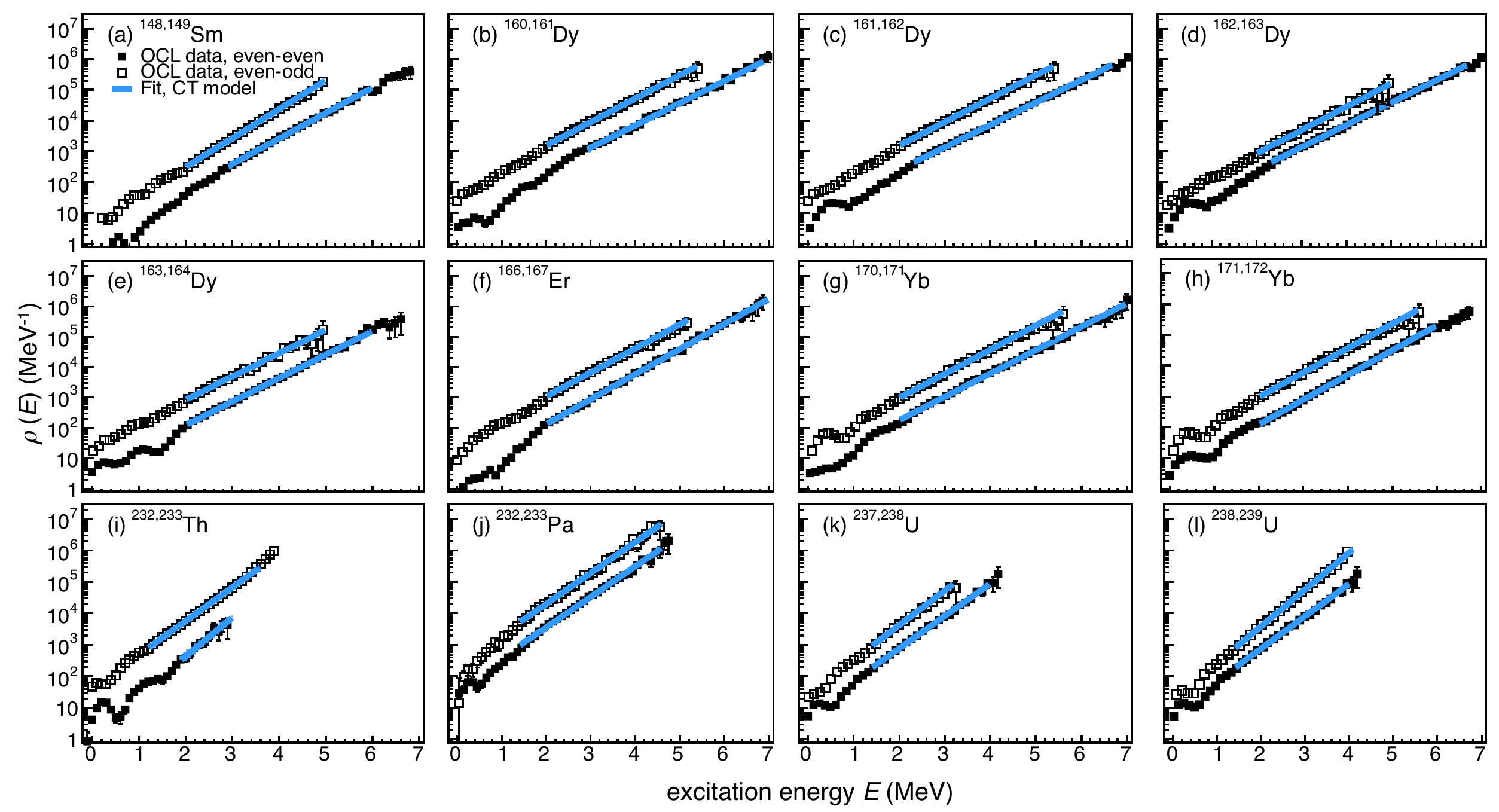}
 \caption {(Color online) Experimental level densities for rare-earth and actinide nuclei measured at 
 the Oslo Cyclotron Laboratory (OCL), 
 with a fit of the 
 constant-temperature model (blue line) for excitation energies above $\approx 2$ MeV: (a) $^{148,149}$Sm~\cite{siem2002}, 
 (b) $^{160,161}$Dy~\cite{guttormsen2003}, (c) $^{161,162}$Dy~\cite{guttormsen2003}, 
 (d) $^{162,163}$Dy~\cite{guttormsen2003,nyhus2012},
 (e) $^{163,164}$Dy~\cite{nyhus2012}, (f) $^{166,167}$Er~\cite{melby2001}, (g) $^{170,171}$Yb~\cite{agvaanluvsan2004},
 (h) $^{171,172}$Yb~\cite{agvaanluvsan2004}, (i) $^{232,233}$Th~\cite{guttormsen2013}, (j) $^{232,233}$Pa~\cite{guttormsen2014},
 (k) $^{237,238}$U~\cite{guttormsen2013} and (l) $^{238,239}$U~\cite{guttormsen2013}.
 }
 \label{fig:nld_rare_earth}
 \end{center}
 \end{figure*}
%---------------------------------------------------%
%---------------------------------------------------%
 \begin{figure}[hbt]
 \begin{center}
 \includegraphics[clip,width=0.9\columnwidth]{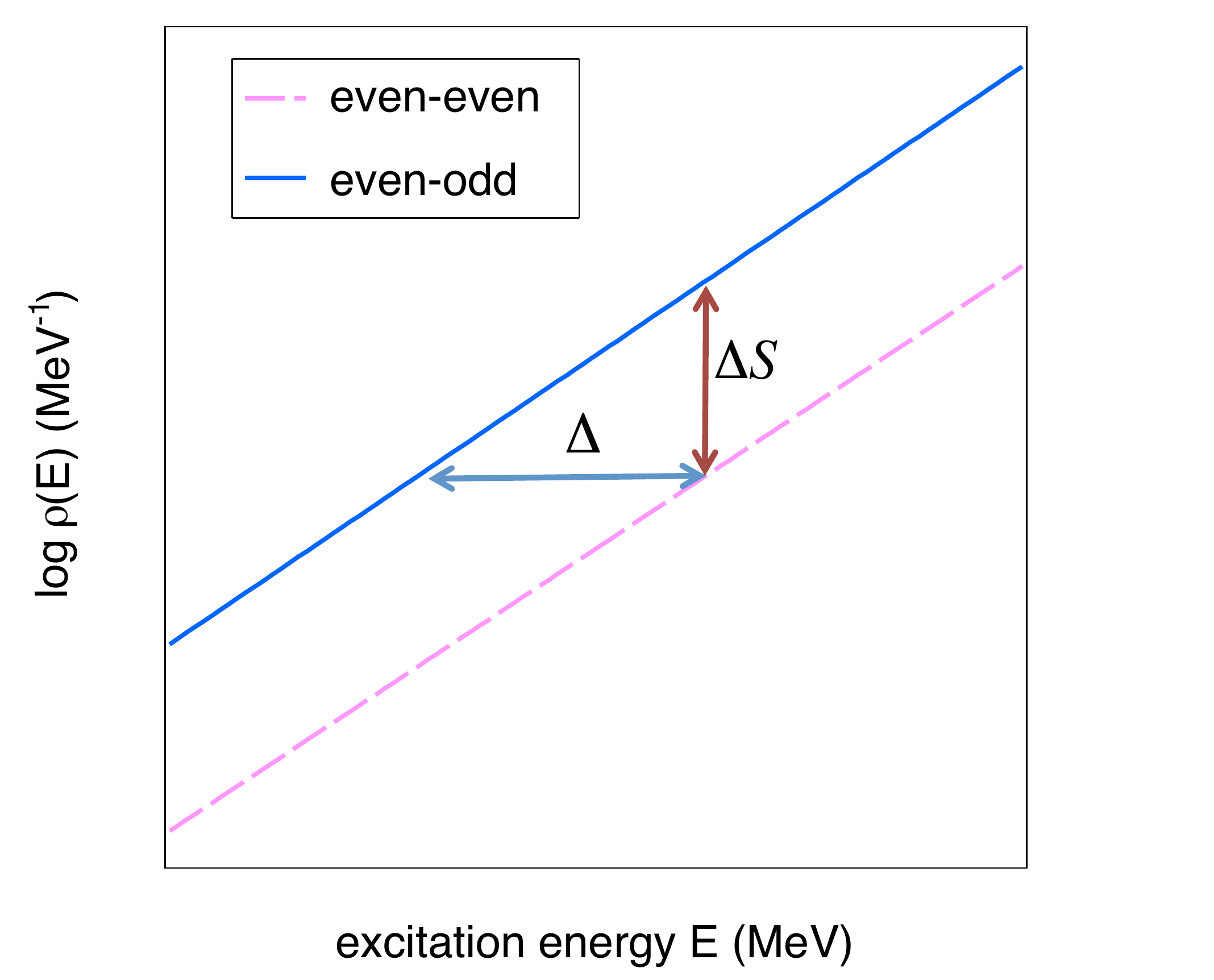}
 \caption {(Color online) Illustration of constant-temperature level densities. The experimental, horizontal shift gives the slope
 ($1/T_{\mathrm{CT}}$) through Eq.~(\ref{eq:delta_BCS}), and the vertical shift is related to the entropy excess
 ($S \sim \ln\rho(E)$, $\Delta S = \ln[\rho_{\mathrm{eo}}(E)/\rho_{\mathrm{ee}}(E)]$)
 for the quasiparticle as indicated in the figure. 
 }
 \label{fig:nld_illustration}
 \end{center}
 \end{figure}
%---------------------------------------------------%

%************************************************************************************%
\begin{table*}[h]
\caption{Extracted temperatures $T_{\mathrm{CT}}$ from fitting the CT-model expression to the level-density 
data of rare-earth and actinide nuclei, %(Figs.~\ref{fig:nld_rare_earth} and~\ref{fig:nld_actinides}),
and the corresponding pair-gap parameters $\Delta_{\mathrm{CT}}$ 
calculated from Eq.~(\ref{eq:delta_BCS}). These are compared to the global formula for $\Delta_{\mathrm{BM}}$ (Eq.~(\ref{eq:BM})),
for which the temperature is deduced using
Eq.~(\ref{eq:delta_BCS}). Also, the even-odd experimental shift, $\rho_{\mathrm{eo}}$ vs. $\rho_{\mathrm{ee}}$, is given,
and the temperature is estimated from this shift for the odd nucleus. The experimental entropy excess is given by 
$\Delta S = \ln[\rho_{\mathrm{eo}}(E)/\rho_{\mathrm{ee}}(E)]$ in units of Boltzmann's constant $k_B$.}
\begin{tabular}{lccccccc}
\hline
\hline
Nuclide    & $T_{\mathrm{BM}}$ & $T_{\mathrm{CT}}$  & $T_{\mathrm{eo}}$  & $\Delta_{\mathrm{BM}}$ & $\Delta_{\mathrm{CT}}$  & $\Delta_{\mathrm{eo}}$ & $\Delta S$ \\
		   & (MeV)    & (MeV)     & (MeV)     & (MeV)         & (MeV)          & (MeV)         & ($k_B$)    \\
\hline
$^{148}$Sm &  $0.56$  & $0.51(1)$ & $-$       & 0.99          & $0.90(2)$      & $-$            &  $-$       \\
$^{149}$Sm &  $0.56$  & $0.46(1)$ & $0.51(6)$ & 0.98          & $0.81(2)$      & $0.9(1)$       &  2.0(2)    \\ % relative to 148Sm
$^{160}$Dy &  $0.54$  & $0.60(1)$ & $-$       & 0.95          & $1.05(1)$      & $-$            &  $-$       \\
$^{161}$Dy &  $0.54$  & $0.58(2)$ & $0.57(6)$ & 0.95          & $1.01(3)$      & $1.0(1)$       &  1.9(2)    \\ % relative to 160Dy
$^{162}$Dy &  $0.54$  & $0.60(1)$ & $-$       & 0.94          & $1.05(2)$      & $-$            &  $-$       \\
$^{163}$Dy &  $0.53$  & $0.57(2)$ & $0.51(6)$ & 0.94          & $1.00(4)$      & $0.9(1)$       &  1.9(2)    \\  % relative to 164Dy
$^{164}$Dy &  $0.53$  & $0.56(1)$ & $-$       & 0.94          & $0.99(1)$      & $-$            &  $-$       \\
$^{166}$Er &  $0.53$  & $0.52(1)$ & $-$       & 0.93          & $0.92(1)$      & $-$            &  $-$       \\
$^{167}$Er &  $0.53$  & $0.56(2)$ & $0.51(6)$ & 0.93          & $0.99(4)$      & $0.9(1)$       &  2.0(2)    \\ % relative to 166Er
$^{170}$Yb &  $0.52$  & $0.57(1)$ & $-$       & 0.92          & $1.00(1)$      & $-$            &  $-$       \\
$^{171}$Yb &  $0.52$  & $0.55(1)$ & $0.51(6)$ & 0.92          & $0.96(1)$      & $0.9(1)$       &  1.8(2)    \\ % relative to 170Yb
$^{172}$Yb &  $0.52$  & $0.54(1)$ & $-$       & 0.91          & $0.95(1)$      & $-$            &  $-$       \\
$^{231}$Th &  $0.45$  & $0.41(1)$ & $0.51(11)$& 0.79          & $0.72(2)$      & $0.9(2)$       &  2.4(4)    \\ % relative to 232Th
$^{232}$Th &  $0.45$  & $0.34(1)$ & $-$       & 0.79          & $0.60(2)$      & $-$            &  $-$       \\
$^{233}$Th &  $0.45$  & $0.40(2)$ & $0.51(11)$& 0.79          & $0.70(2)$      & $0.9(2)$       &  2.3(4)    \\ % relative to 232Th
$^{232}$Pa &  $0.45$  & $0.44(1)$ & $0.40(6)$ & 0.79          & $0.77(2)$      & $0.7(1)$       &  1.7(2)    \\ % relative to 233Pa
$^{233}$Pa &  $0.45$  & $0.45(1)$ & $-$       & 0.79          & $0.79(2)$      & $-$            &  $-$       \\
$^{237}$U  &  $0.44$  & $0.40(1)$ & $0.40(6)$ & 0.78          & $0.70(2)$      & $0.7(1)$       &  1.9(2)    \\ % relative to 238U
$^{238}$U  &  $0.44$  & $0.42(1)$ & $-$       & 0.78          & $0.74(2)$      & $-$            &  $-$       \\
$^{239}$U  &  $0.44$  & $0.37(1)$ & $0.37(3)$ & 0.78          & $0.65(1)$      & $0.65(5)$      &  2.5(5)    \\ % relative to 238U

\hline
\hline
\end{tabular}
\\
\label{tab:deltapar}
\end{table*}
%************************************************************************************%

This experimental linear dependence of the entropy $S(E) \approx \ln \rho(E)$ as given by Eq.~(\ref{eq:ct_nld})
is the microcanonical hallmark of first-order phase transitions. It implies a \textit{constant temperature}, a 
\textit{latent heat}
and an \textit{infinite heat capacity}; 
thus, we may have been staring at the biggest signal yet of such a transition without seeing it. 
Also the experimental data, displayed in Fig.~\ref{fig:nld_rare_earth}
showing data from the rare-earth region~\cite{siem2002,guttormsen2003,nyhus2012,melby2001,agvaanluvsan2004}, 
and several actinides~\cite{guttormsen2013,guttormsen2014}, show that the entropies of adjacent even-even and 
odd-$A$ nuclei are parallel over the 
experimental energy range above $\approx 2$ MeV; i.e., the level densities 
of neighboring even-even and odd-$A$ nuclei have nearly identical slopes,  
Therefore, the level densities of neighboring isotopes can be made to overlap by means of a 
\textit{horizontal shift} along the excitation-energy
axis (see Fig.~\ref{fig:nld_illustration} and Ref.~\cite{guttormsen2000}).
The resulting shift is \textit{constant with energy} and in very good agreement with the
even-odd mass difference; see Table~\ref{tab:deltapar}.
Significantly and surprisingly, this constant shift identifies the elementary excitations as the quasiparticles and 
implies that the energy cost per quasiparticle is \textit{constant} and \textit{independent} of 
excitation energy. 

Equally intriguing is the \textit{vertical shift} between the even-even$-$odd-$A$ nuclear level densities, bringing the lower 
even-even level density on top of the higher odd-$A$ one (see Fig.~\ref{fig:nld_illustration} and Ref.~\cite{guttormsen2001}).
This difference in entropy, approximately constant for excitation energies above $\approx 2$ MeV, 
can be interpreted as the \textit{entropy carried by the extra 
quasiparticle}~\cite{guttormsen2001}.
The experimental evidence alone thus suggests that as the system is excited, quasiparticles are created with 
a \textit{constant energy cost} and carrying a \textit{constant amount of entropy}, see Table~\ref{tab:deltapar}.
This \textit{theory-independent observation} is a clear signature of a first-order phase transition. 
As we shall discuss below, the two phases are a superfluid phase and an ideal gas of quasiparticles. 

In summary, the thermodynamic interpretation of the data is that of a clear first-order phase transition with latent heat
$\Delta$ per (quasi)particle, and infinite heat capacity, and a fixed amount of entropy per (quasi)particle. All of 
the above follows from straightforward thermodynamics, without reference to any specific nuclear-structure theory. 

%%%%%%%%%%%%%%%%%%%%%
% THEORY, BCS       %
%%%%%%%%%%%%%%%%%%%%%

\begin{figure}[hbt]
 \begin{center}
 \includegraphics[clip,width=1.\columnwidth]{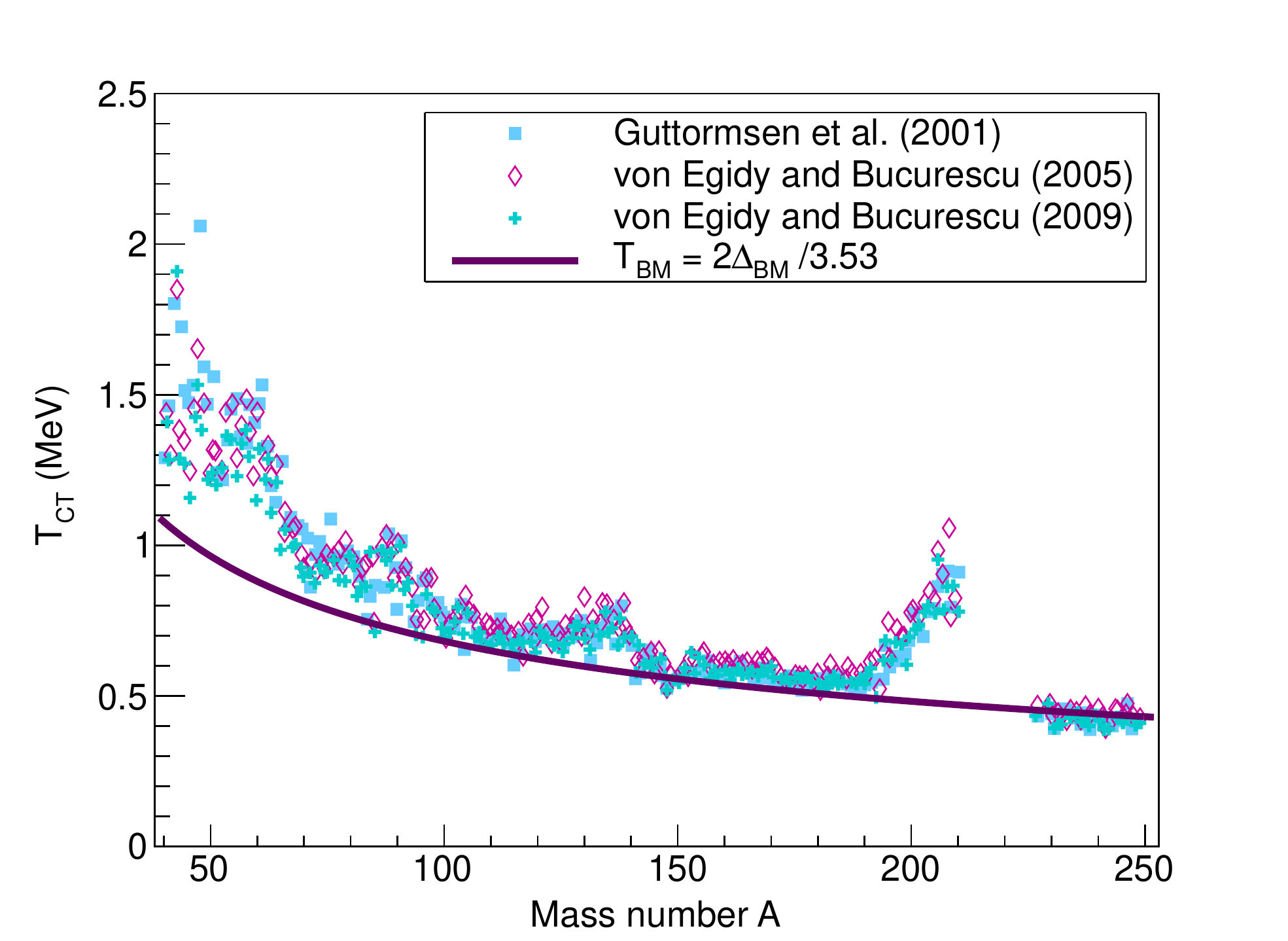}
 \caption {(Color online) Relation between mass number $A$ and constant temperature $T_{\mathrm{CT}}$ taken from 
 Refs.~\cite{guttormsen2001,egidy2005,egidy2009} for $A=40-250$, compared to $T_{\mathrm{cr}}$ from 
 Eq.~(\ref{eq:delta_BCS}) utilizing Eq.~(\ref{eq:BM}) (dark purple line).
 }
 \label{fig:mass_delta_temp}
 \end{center}
 \end{figure}

The phase transition is, at least for nuclei
well away from closed shells, clearly related to pairing. First of all, the constant shift $\Delta$ is directly
related to the liquid drop mass difference, which, in turn, arises from pairing. Furthermore,
if we, provisionally, take the constant 
temperature of the experimental level-density spectrum to be the BCS critical temperature, then, according to the well-known
BCS relation 
\begin{equation}
T_{\mathrm{cr}} = \frac{2\Delta_0}{3.53},
\label{eq:delta_BCS}
\end{equation}
we can extract the gap parameter $\Delta_0$ and compare it directly with that obtained from 
even-odd mass differences represented in the liquid-drop term as described e.g. by 
Bohr and Mottelson~\cite{bohr_mottelson1969}: 
\begin{equation}
\Delta_{\mathrm{BM}} \approx 12A^{-1/2},
\label{eq:BM}
\end{equation}
in units of MeV. For a wide range of mass numbers $A$, the resulting relationship between mass number and temperature using Eq.~(\ref{eq:BM})
is shown in Fig.~\ref{fig:mass_delta_temp},
where the experimental constant temperatures $T_{\mathrm{CT}}$ are taken from Refs.~\cite{guttormsen2001,egidy2005,egidy2009}.
The close agreement in magnitude and trend is remarkable for $A>100$ and away from closed shells, although the assimilation of 
the constant level-density
temperature characteristic of a first-order transition to a critical temperature associated with a second-order 
transition remains to be explained. 

As a consequence of this observation, we can
\textit{predict} from  the even-odd mass difference the low-energy nuclear level densities 
throughout the nuclear chart for regions away from magic proton/neutron numbers. From these features we can immediately
infer that there occurs, at the constant temperature $T_{\mathrm{cr}}$, a first-order phase transition with a latent heat.
%---------------------------------------------------%
%---------------------------------------------------%

%%%%%%%%%%%%%%%%%%%%%
% EXPERIMENTAl DATA %
%%%%%%%%%%%%%%%%%%%%% 
Now we proceed to show that all the empirical features find a close counterpart in BCS theory, 
with due caution for the microcanonical/canonical
language. For a set of uniformly spaced single-particle levels, at the critical temperature the excitation energy 
is given by~\cite{moretto1975}
\begin{equation}
E_{\mathrm{cr}} = \frac{1}{2}g \Delta_0^2 + \frac{\pi^2}{3}gT_{\mathrm{cr}}^2,
\label{eq:Ecritical}
\end{equation}
with $g$ equal to the density of doubly degenerate single-particle states. 
In deformed nuclei the degenerate states are time reversed with spin projections $\Omega$ and $-\Omega$
on the symmetry axis. 
The most probable number of quasiparticles $Q_{\mathrm{cr}}$ at $T_{\mathrm{cr}}$ is~\cite{moretto1975}
\begin{equation}
Q_{\mathrm{cr}} = 4g T_{\mathrm{cr}} \ln 2.
\label{eq:number_qp}
\end{equation}
By taking the ratio of these two quantities and utilizing Eq.~(\ref{eq:delta_BCS}), we get the average cost per created quasiparticle
up to $T_{\mathrm{cr}}$ to be
\begin{equation}
\frac{E_{\mathrm{cr}}}{Q_{\mathrm{cr}}} \simeq \Delta_0.
\label{eq:E_over_Q}
\end{equation}
This is a very puzzling result, though it is consistent with the parallel behavior of the level densities
described above. Within the BCS theory, we know that at $T=0$ the quasiparticle energy 
is $\approx \Delta_0$, and that $\Delta$ decreases with increasing temperature, so that $\Delta=0$ at $T_{\mathrm{cr}}$ 
(see e.g. Fig.~3 of Ref.~\cite{morse1957} for the tin superconductor).

\begin{figure*}[hbt]
 \begin{center}
 \includegraphics[clip,width=1.6\columnwidth]{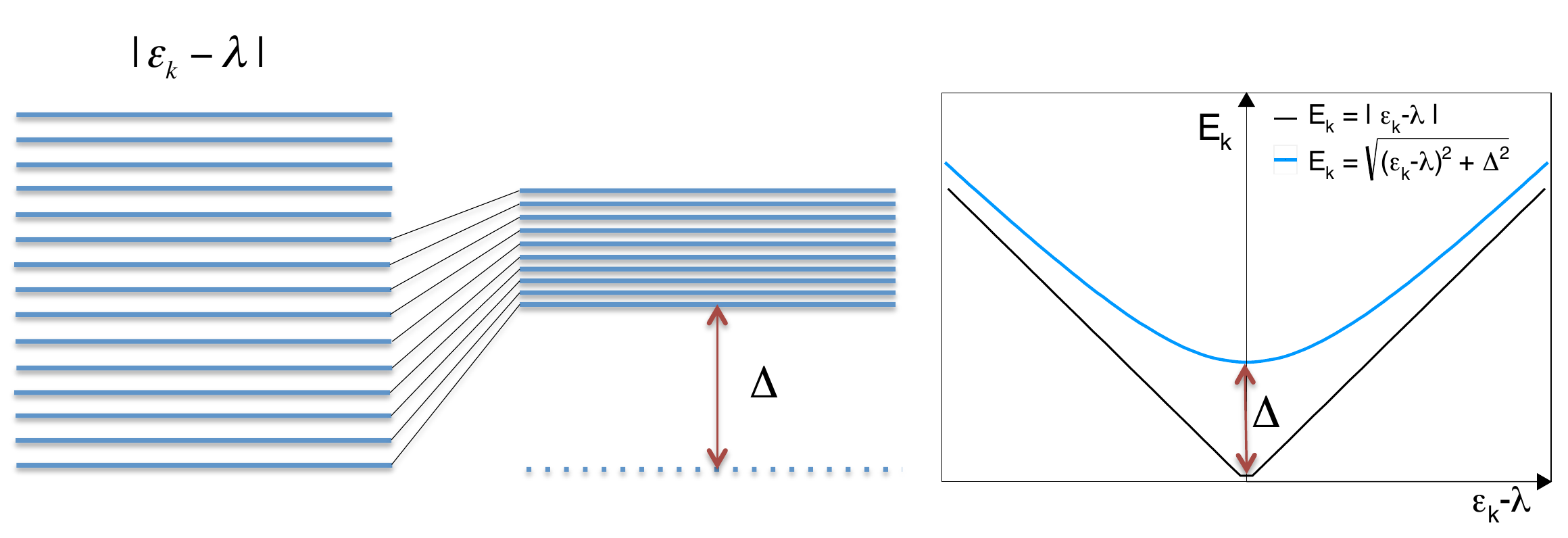}
 \caption {(Color online) Illustration of an equidistant distribution of single-particle levels with energy $|\epsilon_k - \lambda|$,
 where $\lambda$ is the Fermi energy (or, equivalently, the chemical potential) for the case with no pairing (left),
 and when the pairing force comes into play (middle). The quasiparticle energy $E_k$ as a function of $\epsilon_k - \lambda$
 with and without the pairing term ($\Delta^2$) is given in the right part of the figure.}
 \label{fig:singleparticle}
 \end{center}
 \end{figure*}

So, how is it then possible for the energy per quasiparticle to be constant in this excitation-energy range?
The explanation lies mostly in the structure of the quasiparticle energy, which reads
\begin{equation}
E_k = \sqrt{(\epsilon_k-\lambda)^2+\Delta^2},
\label{eq:qp_energy}
\end{equation}
where $\epsilon_k$ and $\lambda$ are the single-particle energy and chemical potential, respectively 
(see Fig.~\ref{fig:singleparticle}).
As the temperature increases, $\Delta$ does indeed decrease, but within the uniform-spacing model this is compensated
by the increase of the average value of $\left|\epsilon_k -\lambda\right|$, and by the change of the underlying
pairing field. 
This fact is immediately seen for the case of a degenerate model, with a single high-degeneracy single-particle level,
for which the ratio of Eq.~(\ref{eq:E_over_Q}) can also be calculated analytically:
\begin{equation}
\frac{E_{\mathrm{cr}}}{Q_{\mathrm{cr}}} = \frac{1}{2}\Delta_0.
\label{eq:E_over_Q_degenerate}
\end{equation}
In this case, the decrease in $\Delta$ with temperature \textit{cannot} be compensated by an increase in the one-body energy 
$\epsilon_k$, as only one degenerate level is available. 
%---------------------------------------------------%
%---------------------------------------------------%

As an example, we have applied the uniform model (pure BCS) with $\Delta_0 = 1$ MeV, and with a doubly-degenerate single-particle 
level density of $g = 7/\mathrm{MeV}$. The parameters would correspond to $A\sim\left(12/1\right)^2 \sim 144$, 
$E_{\mathrm{cr}} = 10.89$ MeV, $Q_{\mathrm{cr}} = 10.99$ and $T_{\mathrm{cr}} = 0.567$ MeV.
For this case, we have calculated the average energy per quasiparticle $E/Q$
as a function of the most probable quasiparticle number in the range $1<Q<Q_{\mathrm{cr}}$. 
%---------------------------------------------------%
 \begin{figure}[tb]
 \begin{center}
 \includegraphics[clip,width=0.9\columnwidth]{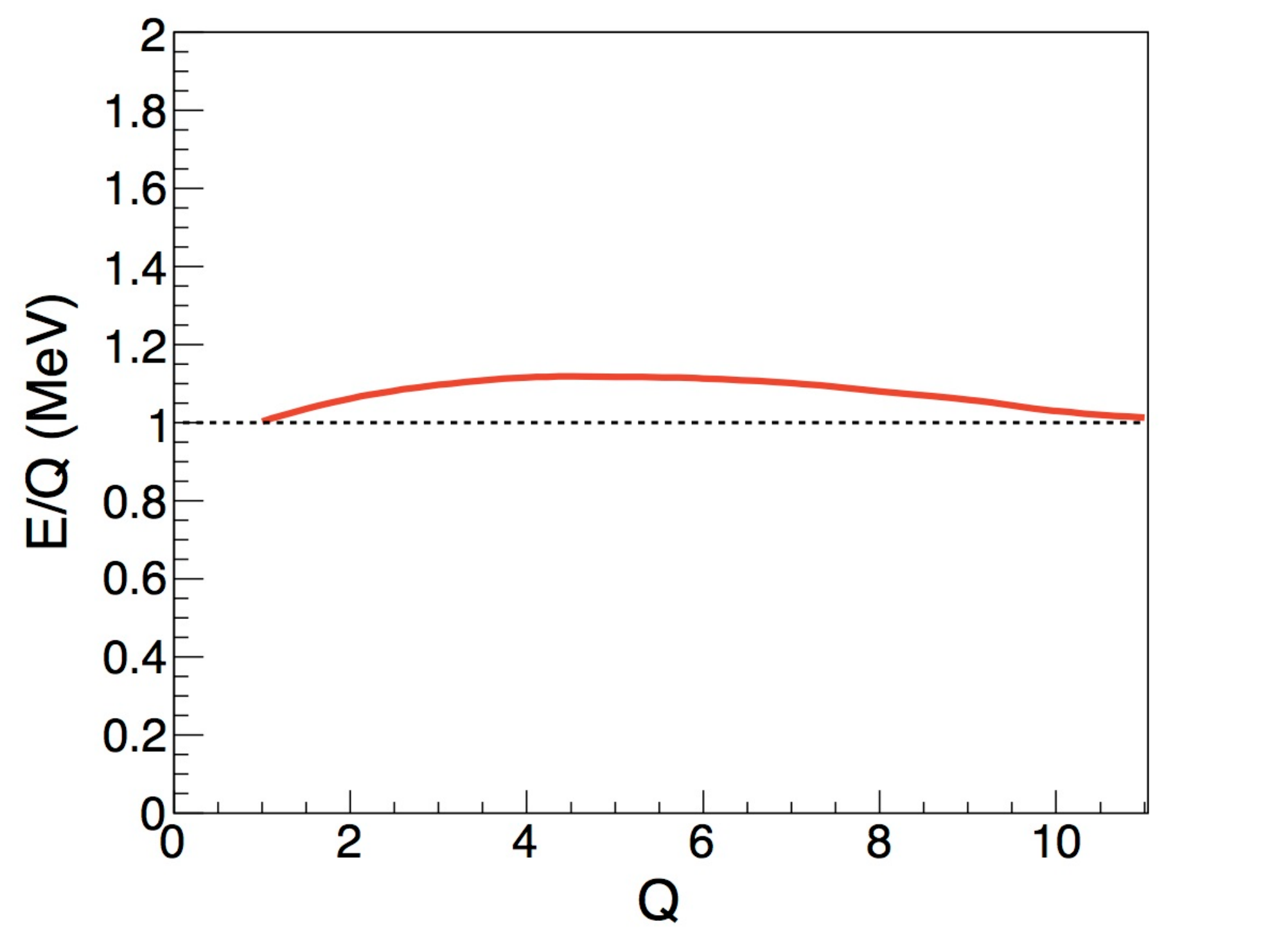}
 \caption {(Color online) Average energy per quasiparticle as a function of the most probable quasiparticle number (see text).}
 \label{fig:probqp}
 \end{center}
 \end{figure}
%---------------------------------------------------%
Figure~\ref{fig:probqp} shows that the energy per quasiparticle is very close to 1 MeV in the entire range. At low $Q$ it increases slightly due
to the one body term in the quasiparticle energy. For higher $Q$ the effect of blocking begins to counteract the one-body effect 
and the ratio decreases slightly until it lands on the analytical result given in Eq.~(\ref{eq:E_over_Q}) at the critical value
$Q_{\mathrm{cr}}$.

The assimilation of $T$ with $T_{\mathrm{cr}}$ finds also an explanation in the BCS model. The dependence of the heat capacity upon
temperature rises exponentially from zero and peaks at $T=T_{\mathrm{cr}}$, so that essentially all energy is absorbed at this temperature.

From the constant energy cost $\Delta$ per quasiparticle, it follows that the entropy per quasiparticle is
\begin{equation}
\frac{\partial S}{\partial Q} = \frac{\Delta_0}{T_{\mathrm{cr}}} = \frac{3.53}{2} = 1.77,
\label{eq:delta_S_perQ}
\end{equation}
to be compared with the empirical, vertical shift as discussed above (see Tab.~\ref{tab:deltapar}).

To summarize, if we use the energy rather than the temperature as the independent variable, 
we observe the progressive creation of 
quasiparticles, in number proportional to the energy, like the amount of ice melted is proportional to the absorbed heat, 
\textit{independent} of the amount of previously melted ice. This independence, together with the constant entropy per quasiparticle, 
gives unmistakable evidence of a first-order phase transition.

The grand-canonical BCS theory parallels almost exactly the experimental findings once we change variable from 
temperature $T$ to \textit{average} energy $\left<E\right>$, except for the presence of a second-order transition.
The experiment, with its microcanonical, non-fluctuating energy $E$ (and particle number $N$) portrays instead a
first-order transition. We have shown how close indeed the BCS treatment, despite its fluctuations, comes to predicting 
the experimentally observed first-order phase transition.

One may wonder whether by transforming the present microcanonical level densities to the canonical ensemble
one could recapture the BCS second-order phase transition. 
So, let us take the microcanonical level density as given in Eq.~(\ref{eq:ct_nld}): 
\begin{equation}
\rho(E) = 1/T_0 \exp(E/T_0),
\end{equation}
and perform the Laplace transform
\begin{eqnarray}
Z(T) &=& \int_0^{\infty}\rho(E)\exp(-E/T) \, dE \nonumber \\
&=& \int_0^{\infty}1/T\exp(-E(1/T - 1/T_0) \,dE \nonumber\\
&=& \frac{T}{T_0-T}.
\end{eqnarray}
Thus, we obtain a partition function that allows all temperatures $0\leq T \leq T_0$, like BCS. We see that for $T=T_0$, we
have a singularity. Is then $T_0$ a limiting temperature, perhaps akin to the BCS $T_{\mathrm{cr}}$? 
However, the microcanonical temperature is $T_0$ and 
only $T_0$ by construction. A well-known physical system with entropy $S = E/T_0$, where $E$ is the heat, is ice/water, 
which at standard condition
has the only temperature $T_0 = 0^\circ$, i.e. a thermostat. The canonical transformation just performed, would lead
to temperatures $0\leq T \leq T_0$, which is in stark contrast with thermodynamics. As is well known, as long as both the
solid and liquid are present, the system can absorb or release energy without any temperature change. So, what is wrong?

For the Laplace transform to be thermodynamically meaningful, it is \textit{necessary} that the entropy 
be a convex function of energy, namely 
\begin{equation}
\partial^2S/\partial^2 E < 0.
\end{equation}
This is not the case here; thus, any result inferred from the partition function $Z$, like heat capacity and other
thermodynamic quantities are incorrect. In conclusion, the only temperature possible for the system is $T_0$, while the
energy at that temperature is undefined. Hence, nuclear systems behave exactly like the water-ice system.

%---------------------------------------------------%
 \begin{figure}[tb]
 \begin{center}
 \includegraphics[clip,width=0.9\columnwidth]{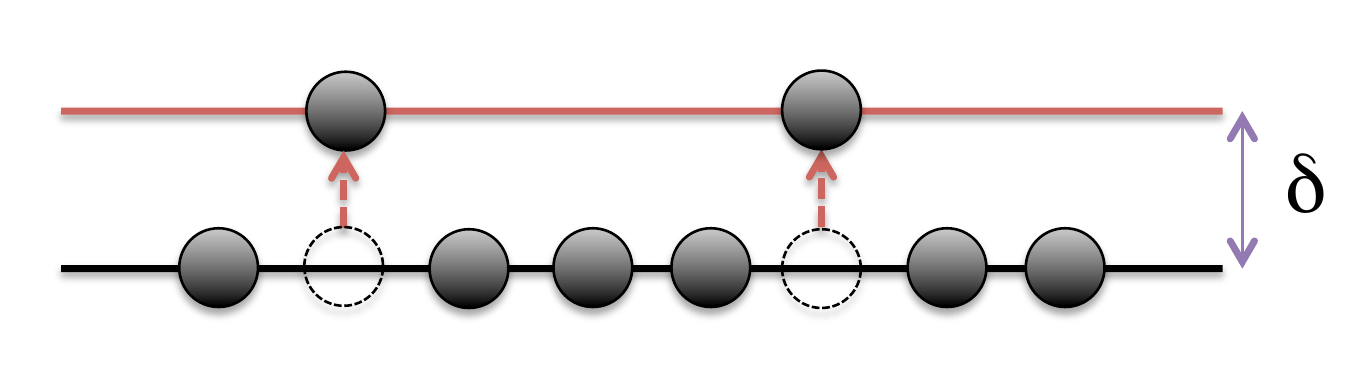}
 \caption {(Color online) Illustration of the case where a gap is present in the single-particle spectrum, 
 but not necessarily due to pairing.
 }
 \label{fig:anygap}
 \end{center}
 \end{figure}
%---------------------------------------------------%
To further illustrate this picture, let us consider a quasiparticle vacuum of degeneracy $N$, 
which can be populated with $n$ quasiparticles at a cost $E = n \delta$, where $\delta$ is the 
energy spacing between the 
vacuum state and the single-quasiparticle level as shown in Fig.~\ref{fig:anygap}. 
The associated number of accessible states is $\omega \approx N^n$, and the entropy reads 
\begin{equation}
S = \ln\omega \approx n \ln N = \frac{E \ln N}{\delta}
\end{equation}
for $n\ll N$, and the entropy per particle is simply
\begin{equation}
\frac{S}{n} = \ln N.
\end{equation}
Thus, the level density is given by
\begin{equation}
\rho(E) \sim \exp S(E) = \exp\left( \frac{E\ln N}{\delta}\right),
\end{equation}
where we easily identify the microcanonical temperature as $T = \delta/\ln N$.
This shows that a low-energy exponential level density is a consequence of a gap in the quasiparticle spectrum
irrespective of its origin, be it due to pairing or shell effects. 
This result can be verified by applying this picture to 
the pairing case, in the context of the BCS model. Let us substitute $\delta$ with $\Delta_0$ and $T$ 
with $T_{\mathrm{cr}}$; utilizing Eq.~(\ref{eq:delta_BCS}), 
we obtain
\begin{equation}
T_{\mathrm{cr}} = \frac{\delta}{\ln N} = \frac{2\Delta_0}{3.53},
\end{equation}
and
\begin{equation}
\ln N =\Delta_0/T_{\mathrm{cr}} = 1.77,
\label{eq:anygap} 
\end{equation}  
in exact agreement with Eq.~(\ref{eq:delta_S_perQ}), and in good agreement with the average, experimental values from 
Table~\ref{tab:deltapar} giving $\Delta S_{\mathrm{ave}}=2.0(1)$ $k_B$.
Further, the number of available states per quasiparticle from Eq.~(\ref{eq:anygap}) is $\exp(\Delta S) = \exp(1.77)\approx 6$,
again agreeing well with the experimental $\Delta S$ values in Table~\ref{tab:deltapar}, $\exp(\Delta S_{\mathrm{ave}})=8(1)$.

The above discussion shows that pairing is just a particular example of a more general case, and that the role of pairing
is just to provide a gap in the quasiparticle spectrum. Returning to the experimental data, we see that 
the vertical entropy shift is essentially constant with excitation energy
and very close in value (see Table~\ref{tab:deltapar}) to the predicted one in Eq.~(\ref{eq:anygap}).
This entropy per quasiparticle reveals the
phase space in which the quasiparticles move as nearly independent objects. The resulting overall picture is that
of a coexistence between an underlying superfluid phase in equilibrium with an ideal gas of almost independent
quasiparticles. This picture comes about mostly, but not wholly, by a shift of perspective, from the canonical to microcanonical
approach or from temperature to energy as the independent variable. Also, the experimental availability of entropy versus energy for 
even-even and odd-$A$ or odd-odd nuclei allows one to observe the very pictorial
feature of gap superconductors and gap-less superconductors in mesoscopic systems. 
We expect these features and their interpretation to be very relevant for other mesoscopic systems, 
such as superconductive clusters differing by just one electron.

In conclusion, we have shown that the experimental low-energy level densities (below the neutron separation energy) portray a strong
signature of a first-order phase transition, which can be understood within the BCS framework. The coexistence of a superfluid
with a gas of quasiparticles is easily characterized thermodynamically, especially through the comparison of even-even
and odd-$A$ nuclei. In particular, it is shown that the even-odd mass difference is sufficient to determine the level density 
in absolute value and energy dependence. The difference observed (and predicted) between even-even and odd nuclei nicely
displays the behavior of gap and gapless superconductors and is also of relevance for superconducting clusters. 
The generalization to systems characterized by any gap, irrespective 
of its origin, allows us to interpret level densities also for nuclei near the magic shells.

\section*{Acknowledgements}
The authors gratefully acknowledge funding from the Research Council of Norway, project grant no. 205528 (A.~C.~L)
and 210007 (F.~G. and S.~S.), and support from the RCN Leiv Eiriksson mobility programme (L.~G.~M.).

\end{document}